# A comparison of possibilities to measure the coordinates of a moving hot body using an infrared telescope or Doppler radar

*S. N. Dolya*

*Joint Institute for Nuclear Research, Joliot Curie str. 6, Dubna, Russia, 141980*

**Abstract**

A comparison of the sensitivities of methods which allow us to determine the coordinates of a moving hot body is made. The infrared telescope can reliably distinguish the signal from the body against the background of thermal radiation from the Earth's surface and reflected solar radiation. The accuracy of determination of the body's coordinates is of the order of 100 meters. The Doppler radar has an accuracy of coordinates' determination of about 10 kilometers and requires deployment in the Earth's orbit of a greater number of satellites compared to the infrared telescope.

**Introduction**

The infrared telescope and the Doppler radar are research instruments with a very high sensitivity. Three telescopes or radars of such type that are placed at the vertices of a triangle in which the Earth is "inscribed" will observe its total circumference. In order to view the entire surface of the Earth, three additional satellites rotating in the orthogonal plane ought to be used.

Let us assume that a hot body moves above the Earth's surface. We will compare then the sensitivities of the above two instruments in determining the body's coordinates.

For that purpose we will calculate the intensity of radiation received by the antenna (mirror) with a diameter of 4 m which is placed at a satellite rotating round a circular orbit at the distance $H_{sat}$ = 6400 km from the Earth's surface. Such a satellite can view a region of order of about $10^4 * 10^4$ km$^2$ on the surface of the Earth.

Three satellites rotating in an equatorial orbit will view a ring with a width from the 0th to 60th parallel north and south, i.e. approximately from the South Polar Circle (67° of south latitude) to the Arctic Circle (67$^0$ of north latitude).

Three more satellites which move along a meridional orbit will observe then the polar regions. And a satellite above the pole will survey the area from the pole to the 30th parallel, which is almost up to the tropic located at the 23rd parallel.

**I. Infrared telescope**

*1. Sensitivity of instruments*

Planck's law gives the radiation intensity of a blackbody as a function of the temperature. A curve for the spectral intensity of the radiation flux against the body's temperature is depicted in [1]. It can be determined from the curve that for the temperature $T = 800 \text{ K}^0$ the radiation power equals:

$$W_{infrared} = 10^{-1} \text{ (W* str}^{-1} \text{ cm}^{-2} \text{ } \mu^{-1}). \quad (1)$$

The radiation peak is then in the wavelength range of 3-4 microns. Shown in [1] is also a curve of sensitivity of various infrared detectors. From this curve it follows that the InSb (-196 $C^0$) detector is suitable one for this wavelength region, and its sensitivity here is: $D^* = 10^{11}$ (cm * Hz$^{1/2}$/W).

We define now the geometric factor G of the receiving instruments as the ratio of the IR telescope (12.56 m$^2$) to the square of the radius. Then $G = 12.56 \text{ m}^2/4\pi H_{sat}^2 = 2.4 * 10^{-14}$. The value $4\pi G$, which is the spatial angle occupied by the telescope mirror, is: $4\pi * 2.4 * 10^{-14} = 3 * 10^{-13}$ str.

We assume that the radiator's area $S_{rad} = 3 * 10^4$ cm$^2$. Then the power coming from the body to the telescope mirror is:

$$W_{infrared} = 10^{-1}*4\pi *2.4*10^{-14}*3*10^4 = 9*10^{-10} \text{ W}.$$

The matrix of the above photo detector must contain $10^5 * 10^5$ pixels and occupy an area of 1m * 1m. Here, each pixel corresponds to a region on the Earth's surface with dimensions 100 m * 100 m.

We determine now the required bandwidth of this detector. A satellite will travel with a speed of 10 km /s, passing through one pixel in the matrix during $10^{-2}$ s. The reciprocal of $10^{-2}$ s, i.e. $10^2$ Hz, is the required bandwidth of a receiving photo detector in the matrix. The expression for the sensitivity of this detector will contain this value as a square root, that is the numerical value of this quantity is 10 Hz$^{1/2}$.

Let us calculate next the area of one photo detector in the matrix. The entire matrix contains about $10^5 * 10^5$ pixels and occupies an area of 1 * 1 m. Then the size of one pixel is 7 * 7 µ.

One cannot use here a photo detector of smaller size because radiation cannot be focused into a region smaller than the wavelength. In our case, the largest wavelength of radiation is: $\lambda = 4$ µ. So, we choose the size of one pixel as $7 * 10^{-4} * 7 * 10^{-4}$ cm$^2$.

The expression for the sensitivity of the detector includes this value also as a square root of the size. Therefore, the actual sensitivity of the detector, $1/D^*$, is: $1/D^* = 10^{-11}$ [W / (cm * Hz$^{1/2}$)] $= 10^{-11} * 7*10^{-4} * 10 = 7 * 10^{-14}$ W.

This is the threshold sensitivity of the detector, $1/D^* = 7 * 10^{-14}$ W. It is four orders of magnitude smaller than the value of the signal: $W_{infrared} = 9 * 10^{-10}$ W. Thus, it can be seen that the detector sensitivity is sufficient as the radiation is focused here to one pixel.

*2. Background conditions. Thermal radiation from the Earth's surface*

Each pixel in our case corresponds to a region of size 100 * 100 m where a hot radiating body is located. This body is "hot", with T = 800 K$^0$, but its surface area is only 3 m$^2$. Next, we assume that the Earth's surface over which this body is found has a temperature of 273 K$^0$ (0 degrees Celsius), and its area is 100 m$^2$.

The intensity of radiation emitted by a body with T = 273 K$^0$ in the wavelength range 3-4 micron is $10^{-5}$ (W / str * cm$^2$ * µ), i.e. four orders of magnitude lower than the radiation from a body with T = 800 K$^0$. The ratio of the surface areas of the body and the Earth is 3 m$^2$/$10^4$ m$^2$ = $3 * 10^{-4}$.

This means that the signal here will be three times greater than the background.

*3. Background conditions. Reflected solar radiation*

The power of solar radiation near the Earth's surface is of order 1 kW/m$^2$. We assume that such power is uniformly distributed over the spectral range with a width of 100 µ. This denotes that the spectral range of width 1 µ (from 3 to 4 µ) will account for an incident power of 10 W/m$^2$.

Let us take the coefficient of reflection of solar radiation from the Earth's surface as 30%. We obtain then that the intensity of reflected radiation falling on one steradian is: $P_{refl} = 10 * 10^4 * 0.3/4\pi = 2.4 * 10^3 \text{ W} * \text{str}^{-1}$.

The power $P_{refl} = 2.4 * 10^3 \text{ W} * \text{str}^{-1}$ is of the same order as the radiation power of a hot body, $W_{infrared} = 10^{-1} * 3 * 10^4 = 3 * 10^3 \text{ W} * \text{str}^{-1}$. This suggests that in the areas with direct incident and reflected solar radiation, the operation of an infrared telescope will be inefficient. Such a telescope can be effective only where the solar radiation is slanting or absent.

**II. Doppler radar**

We will now calculate the intensity of reflected microwave radiation falling on a spherical mirror with a diameter of 4 m which is placed on a satellite of the same kind.

*1. Equipment*

Let us choose a wavelength for the radiation that will be emitted onto the Earth's surface from a satellite: $\lambda_{irr} = 0.4$ cm, which corresponds to the wave frequency $f_{rad} = c / \lambda_{irr} = 7.5 * 10^{10}$ Hz.

We assumes that the first coefficient $G_1$ in the ratio between the power of the incident and reflected waves equals:

$$G_1 = 0.5. \qquad (2)$$

The second factor is related to the aperture of the receiving apparatus: $G_2 = S_{ant}/4\pi H_{sat}^2$. The geometric factor $G_2$ is thus:

$$G_2 = S_{ant}/4\pi H_{sat}^2 = 2.4*10^{-14}. \qquad (3)$$

A satellite can view a region of about $10^4 * 10^4$ km$^2$. The area of such a visible region is: $S_{reg} \approx 2 * 10^8$ km$^2$. Assume that the satellite observes one by one smaller plots with the dimensions 10 x 10 km$^2$. Then the area of each plot $S_{plot} = 100$ km$^2$ and the number of such viewed plots $S_{reg} / S_{plot} = 2 * 10^6$.

Assume also that the surface area of the body moving above the Earth's surface $S_{body} = 1$ m$^2$, then the ratio of the body's surface area to the area of the viewed plot gives another factor: $G_3 = S_{body} / S_{plot}$. In our case, this ratio is:

$$G_3 = S_{body} / S_{plot} = 10^{-8}. \qquad (4)$$

Let the body's velocity $V_{body}$ be equal to 0.3 - 3 km / s. The Doppler frequency shift $\Delta f_{Dop} / f_{rad} = V_{body} / c = 10^{-6} - 10^{-5}$, where $c = 3 * 10^{10}$ cm / s is the speed of light in vacuum. In absolute values such frequency shift $\Delta f_{Dop}$ for the selected parameters equals 75 - 750 kHz. It can be easily detected by various methods.

We consider now a sequence of 100 superconducting cavities [2] with a Q-factor of $10^7$ and a bandwidth $\Delta f_{cav}$ of 7.5 kHz. In order to find out whether there is a signal with a frequency corresponding to shifted one and which frequency has the incoming signal, each of the 100 cavities needs to be excited by a signal. A signal due to one cavity will be 100 times smaller than the whole incoming signal. The coefficient $G_4$ in this case is as follows:

$$G_4 = 10^{-2}. \qquad (5)$$

By multiplying all of the above coefficients, we obtain:

$$G_1 * G_2 * G_3 * G_4 = 0.5 * 2.4 * 10^{-14} * 10^{-8} * 10^{-2} = 10^{-24}. \qquad (6)$$

This denotes that if the incident power $P_{rad} = 1$ kW, then the received power $P_{rec} = 10^{-21}$ W. Such power can be detected with modern equipment.

It is assumed that a satellite observes all of the plots during the time $\tau_{obs} = 3$ s, so that the time $\tau_{obs1} = 1.5$ μs is spent for viewing one plot. For a superconducting cavity [2], the intensity of oscillations associated with the thermal noise is following:

$$P_{noi} = kT\Delta f_{cav}, \qquad (7)$$

where $k = 1.38 * 10^{-23}$ J / degree is the Boltzmann constant; T is the cavity's temperature in Kelvin degrees; $\Delta f_{cav} = 7.5$ kHz is the frequency bandwidth of the cavity. For the temperature $T = 1.4$ $K^0$ [2], the cavity's noise power is:

$$P_{noi} = kT\Delta f_{cav} = 1.5 * 10^{-19} \text{ W}. \qquad (8)$$

*2. Accumulation*

When exciting a high-Q cavity by a harmonic signal, the square of the amplitude of its oscillations increases linearly up to values of the order $A^2_{sign} \sim P_{rad} * Q$. During $\tau \approx Q / f_{rad} = 130$ μs the signal grows up to a maximum value. In our case, the time of irradiation and viewing one plot is 1.5 μs, i.e. two orders of magnitude smaller.

For the selected parameters the square of the amplitude of oscillations in the cavity will correspond to the signal power:

$$P_{rad}*Q = 10^{-21}*10^{7}*1.5*10^{-6}/(130*10^{-6}) = 10^{-16} \text{ W}, \qquad (9)$$

which is almost three orders of magnitude greater than the square of the amplitude corresponding to the cavity's noise power $P_{noi} = 1.5 * 10^{-19}$ W.

**Conclusion**

From the above it is evident that the coordinates of a moving hot body can be registered reliably both by an infrared telescope and a Doppler radar.

The frequency shift of the reflected signal due to the Doppler effect is proportional to the projection of velocity directed toward or away from the satellite. To avoid dead zones and improve the sensitivity of the method, it is required, obviously, to increase the number of satellites. Such satellites must have different inclinations of their orbits to the Earth's axis.

References

1. http://www.hamamatsu.com/resources/pdf/ssd/infrared_techinfo_e.pdf

2. http://www.linearcollider.org/about/Publications/Reference-Design-Report, http://en.wikipedia.org/wiki/Superconducting_radio_frequency